\documentclass[10pt]{article}

\usepackage[margin=1in]{geometry}
\usepackage{amsmath}
\usepackage{amssymb}
\usepackage{amsthm}
\usepackage{graphicx}
\usepackage{color}
\usepackage{hyperref}
\usepackage{listings}
\usepackage{siunitx}
\usepackage{caption}
\usepackage{authblk}
\usepackage{cite}

\usepackage{xcolor}
\definecolor{LightGray}{gray}{0.95}
\usepackage[frozencache=true,cachedir=minted-cache]{minted}

\newcommand{\OO}[1]{\mathcal{O}\paren{#1}}
\newcommand{\paren}[1]{\left( #1 \right)}
\newcommand{\brak}[1]{\left[ #1 \right]}

\newcommand{\wh}[1]{\widehat{#1}}

\newcommand{\abs}[1]{\left| #1 \right|}

\newcommand{\omax}{\omega_{\max}}

\newcommand{\gdlr}{G_{\text{DLR}}}

\DeclareMathOperator\supp{supp}

\title{libdlr: Efficient imaginary time calculations using \\ the discrete Lehmann representation}

\author[1,2]{Jason Kaye \thanks{jkaye@flatironinstitute.org}}
\author[2]{Kun Chen \thanks{kunchen@flatironinstitute.org}}
\author[3]{Hugo U. R. Strand \thanks{hugo.strand@oru.se}}

\affil[1]{{\footnotesize Center for Computational Mathematics, Flatiron Institute, New York, NY 10010, USA}}
\affil[2]{{\footnotesize Center for Computational Quantum Physics, Flatiron Institute, New York, NY 10010, USA}}
\affil[3]{{\footnotesize School of Science and Technology, \"Orebro University, Fakultetsgatan 1, SE-701 82, \"Orebro, Sweden}}

\date{}

\begin{document}

\maketitle

\begin{abstract}

  We introduce \texttt{libdlr}, a library implementing the recently
  introduced discrete Lehmann representation (DLR) of imaginary time
  Green's functions.  The DLR basis consists of a collection of
  exponentials chosen by the interpolative decomposition to ensure
  stable and efficient recovery of Green's functions from imaginary time
  or Matsubara frequency samples.  The library provides subroutines to
  build the DLR basis and grids, and to carry out various standard
  operations. The simplicity of the DLR makes it straightforward to
  incorporate into existing codes as a replacement for less efficient
  representations of imaginary time Green's functions, and
  \texttt{libdlr} is intended to facilitate this process.
  \texttt{libdlr} is written in Fortran,
  provides a C header interface,
  and contains a Python module \texttt{pydlr}.
  We also introduce a stand-alone Julia implementation,
  \texttt{Lehmann.jl}.

\end{abstract}

\section{Introduction}

Imaginary time Green's functions are a fundamental component of quantum
many-body calculations at finite temperature. Since the cost of many
algorithms scales with the number of imaginary time degrees of freedom,
there has been significant recent interest in developing efficient
methods of discretizing the imaginary time domain.

The most common approach used in scientific codes is to discretize
imaginary time Green's functions on a uniform grid, or equivalently by a
Fourier series. Although it is simple to use, this representation
requires $\OO{\Lambda/\epsilon}$ degrees of freedom, where $\epsilon$ is
the desired accuracy and $\Lambda = \beta \omega_{\max}$ is a
dimensionless energy cutoff depending on the inverse temperature $\beta$
and a real frequency cutoff $\omega_{\max}$. This scaling makes
calculations at low temperature and high accuracy prohibitively
expensive.
Representations of Green's
functions by orthogonal polynomials require $\OO{\sqrt{\Lambda}
\log(1/\epsilon)}$ degrees of freedom, largely addressing the accuracy
issue but remaining suboptimal in the low temperature limit
\cite{boehnke11,gull18,dong20}.  This has motivated the development of
optimized basis sets in which to expand imaginary time Green's functions;
namely the intermediate representation (IR)
\cite{shinaoka17,chikano18,chikano19} with sparse sampling \cite{li20},
and the more recently introduced discrete Lehmann representation (DLR)
\cite{kaye21}.  Both methods require only
$\OO{\log(\Lambda) \log(1/\epsilon)}$ degrees of freedom, enabling
accurate and efficient calculations at very low temperature. 
The library \texttt{libdlr} implements the DLR, and is intended to
provide the same ease of use as standard uniform grid/Fourier series
discretizations.

Both the IR and the DLR are derived from the spectral Lehmann
representation of imaginary time Green's functions, given by \cite{jarrell96}
\begin{equation} \label{eq:lehmann}
  G(\tau) = -\int_{-\infty}^{\infty} K(\tau,\omega) \rho(\omega) \, d\omega
\end{equation}
for $\tau \in [0,\beta]$, with
\begin{equation} \label{eq:defK}
  K(\tau,\omega) =  \frac{e^{-\omega \tau}}{1+e^{-\beta \omega}}
\end{equation}
and $\rho$ an integrable spectral density. The main observation underlying these
representations is that the integral operator in \eqref{eq:lehmann},
with limits of integration suitably truncated to the support
$[-\omax,\omax]$ of $\rho$,
is well approximated by a low rank operator, so that its image is well
represented by a compact basis. The IR
uses the singular value decomposition of a suitable discretization
of the operator, and the IR basis is constructed from the left singular
vectors. The DLR instead uses the interpolative decomposition (ID)
\cite{cheng05,liberty07}, and the DLR basis is given explicitly by the
functions $K(\tau,\omega_l)$ for a
representative set of frequencies $\{\omega_l\}_{l=1}^r$. In other
words, any imaginary time Green's function obeying a given energy cutoff
$\Lambda$
has a representation
\begin{equation} \label{eq:dlrapprox}
  G(\tau) \approx \gdlr(\tau) = \sum_{l=1}^r K(\tau,\omega_l) \wh{g}_l,
\end{equation}
for some coefficients $\wh{g}_l$, accurate to a user-provided tolerance
$\epsilon$. Whereas the
IR basis is orthogonal and non-explicit, the DLR basis
is non-orthogonal and explicit---the basis functions are simply
exponentials. We emphasize that the frequencies
$\omega_l$ depend only on $\Lambda$ and $\epsilon$, and not on a particular
Green's function.

The simple form of the DLR basis makes it easy to work with.
Standard operations, including transformation to the Matsubara
frequency domain, convolution, and integration, can be carried out explicitly.
Compact grids can be constructed, in both the imaginary time and
Matsubara frequency domains, so that the DLR coefficients $\wh{g}_l$ of
a given Green's function $G$ can be recovered in a stable manner from the values of $G$ at
the grid points. DLR expansions can be multiplied, either in imaginary
time or Matsubara frequency, by simply multiplying their values on the
corresponding grids. Furthermore, algorithms are available to build the DLR basis and grids at a cost which is typically
negligible even for extremely low temperature calculations, and with
controllable, user-determined accuracy guarantees. It is therefore
straightforward to replace less efficient discretizations of imaginary
time Green's functions by the DLR.

The library \texttt{libdlr} provides routines to build the DLR and
associated grids, and to carry out basic operations involving imaginary
time Green's functions.
The library is implemented in Fortran and its only external dependencies
are BLAS and LAPACK. It is straightforward to use on its own,
or to incorporate into existing codes.
The library includes a Python module, \texttt{pydlr}, which can be used
independently or as a wrapper to call \texttt{libdlr}. We have also
built a stand-alone Julia implementation, \texttt{Lehmann.jl}, which
provides similar functionality.

This paper is organized as follows. In Section \ref{sec:dlr}, we provide
a brief overview of the DLR. Section \ref{sec:features} 
describes the main features of \texttt{libdlr}. In Section
\ref{sec:examples}, we give a few Fortran code examples, demonstrating
recovery of a DLR from samples of a Green's function, and the
efficient self-consistent solution of the Sachdev-Ye-Kitaev model. A concluding
discussion is given in Section \ref{sec:conclusion}. Appendices
\ref{app:python} and \ref{app:julia} discuss \texttt{pydlr} and
\texttt{Lehmann.jl}, respectively, with several code examples. Appendix \ref{app:rel} contains a technical
discussion on the relative format used to represent imaginary time
points for high accuracy calculations in \texttt{libdlr}.

\section{Discrete Lehmann representation} \label{sec:dlr}

We begin with a brief overview of the DLR, following \cite{kaye21}, where more details can be found. The
derivation starts with the Lehmann representation
\eqref{eq:lehmann}. We assume the support of the spectral density $\rho$
is contained in $[-\omax,\omax]$, and transform to the dimensionless
variables $\tau \gets \tau/\beta$ and $\omega \gets \beta \omega$.
In these variables, $G$ satisfies a truncated Lehmann representation
\begin{equation} \label{eq:tlehmann}
  G(\tau) = -\int_{-\Lambda}^{\Lambda} K(\tau,\omega) \rho(\omega) \, d\omega,
\end{equation}
for $K$ given by \eqref{eq:defK} with $\beta = 1$, $\tau \in [0,1]$, and
$\Lambda = \beta \omax$.
It has been observed that the singular values of the integral operator
in \eqref{eq:tlehmann} decay super-exponentially
\cite{shinaoka17,kaye21}, suggesting the possibility of
approximating it to accuracy $\epsilon$ by a low rank operator.

In particular, the DLR makes use of a low rank decomposition of $K$ of
the form
\begin{equation} \label{eq:klowrank}
  K(\tau,\omega) \approx \sum_{l=1}^r K(\tau,\omega_l) \pi_l(\omega),
\end{equation}
for smooth functions $\pi_l$. Substitution of such a decomposition into
\eqref{eq:tlehmann} yields the DLR \eqref{eq:dlrapprox}, demonstrating
its existence in principle. As we will see, in practice, we do not obtain the DLR
expansion of a Green's function in this manner, since the spectral function
$\rho$ is typically not known.

\begin{figure}[t]
  \centering
  \begin{minipage}[t]{0.25\textwidth}
    \includegraphics[width=\textwidth]{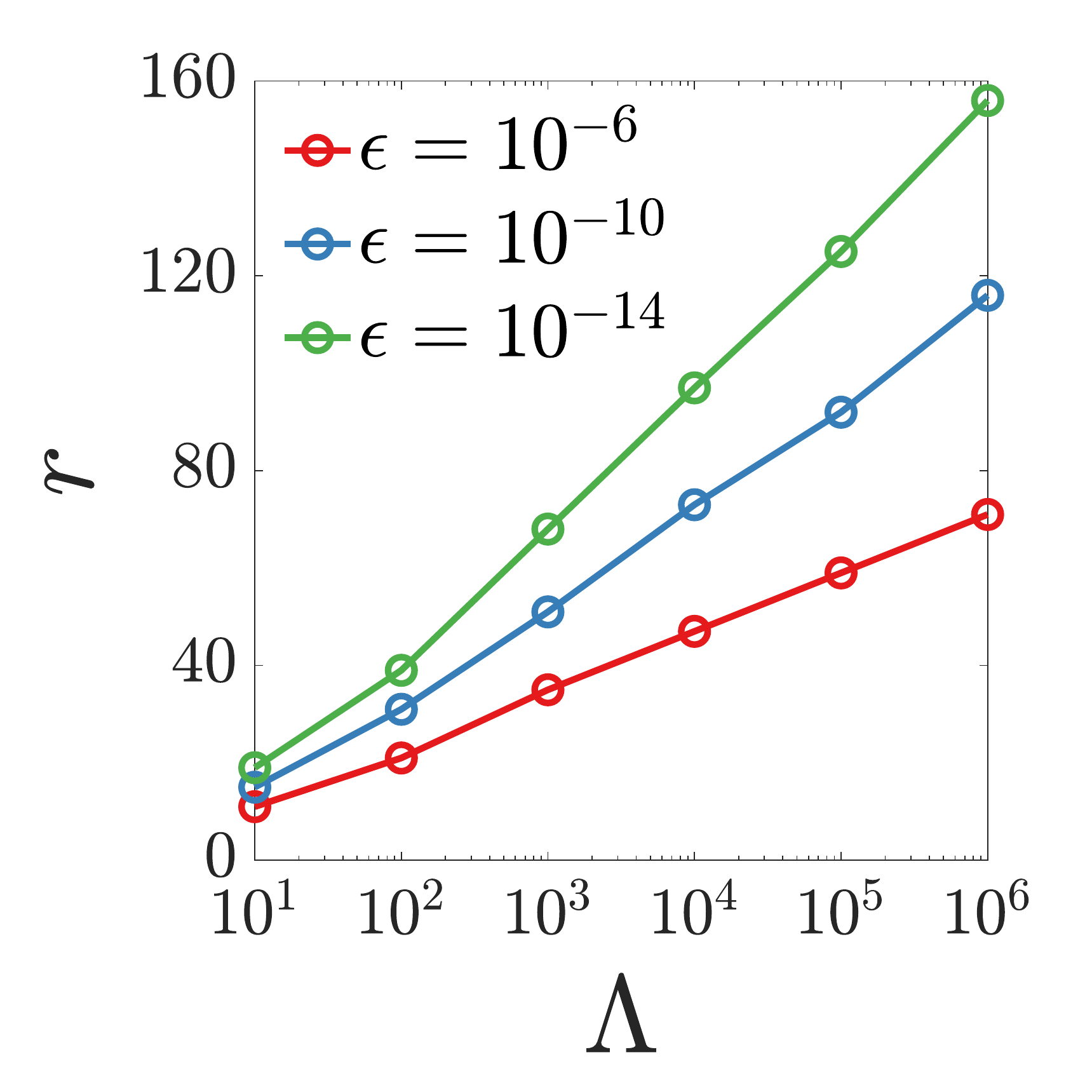}
    \caption{Number $r$ of DLR basis functions for different values of
    $\Lambda$ and $\epsilon$; observed scaling is $r = \OO{\log
    \paren{\Lambda} \log \paren{1/\epsilon}}$. This figure uses data from \cite{kaye21}.}
    \label{fig:dlrrank}
  \end{minipage}
  \hfill
  \begin{minipage}[t]{0.7\textwidth}
    \includegraphics[width=\textwidth]{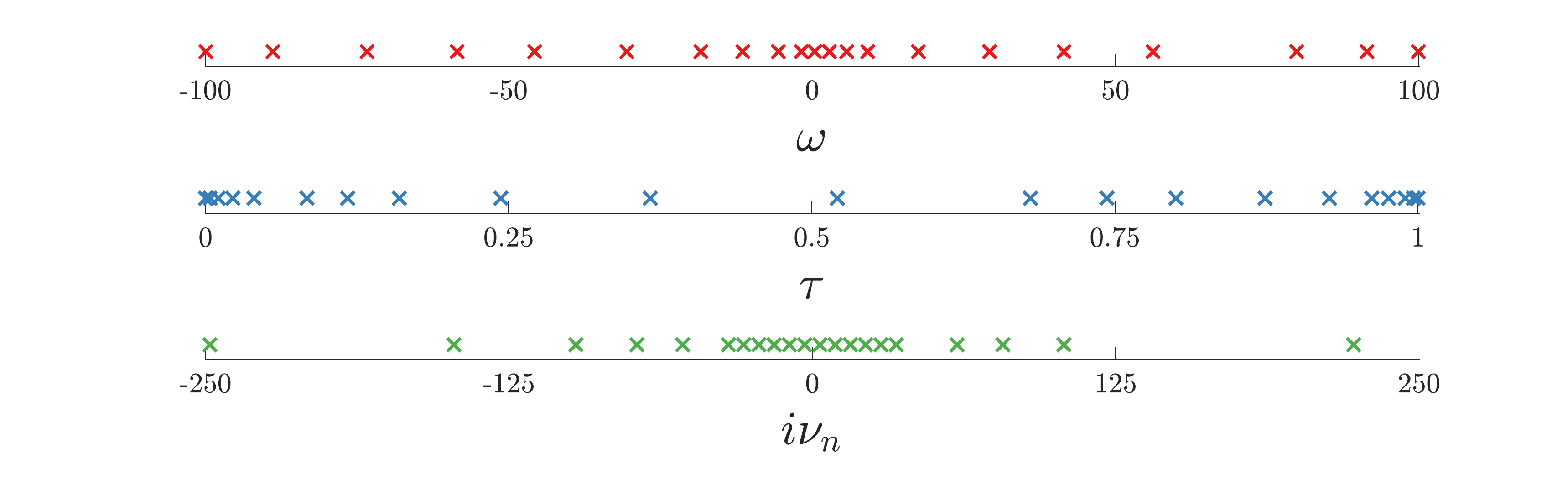}
    \caption{The 21 DLR frequencies, imaginary time nodes, and Matsubara
    frequency nodes for $\Lambda = 100$ and $\epsilon = 10^{-6}$. The
    clustering of each set of nodes reflects the structure of Green's
    functions in the respective domains, and is automatically determined
    by the pivoted QR procedure.}
    \label{fig:nodes}
  \end{minipage}
\end{figure}

The DLR basis $\{K(\tau,\omega_l)\}_{l=1}^r$ is characterized by the DLR frequencies
$\{\omega_l\}_{l=1}^r$, which depend only on the high energy cutoff
$\Lambda$ and the desired accuracy $\epsilon$, both
user-defined parameters.
The DLR frequencies are obtained by a two-step procedure.
First, the kernel $K(\tau,\omega)$ is discretized on a high-order,
adaptive fine grid
$(\tau_i^f,\omega_j^f)_{i,j=1}^{M,N} \subset [0,1] \times
[-\Lambda,\Lambda]$, constructed to efficiently
resolve small scale features. This yields an accurate discretization $A_{ij} =
K(\tau_i^f,\omega_j^f)$ of the integral operator in \eqref{eq:tlehmann}.
Then, a rank revealing column pivoted QR algorithm
is applied to $A$, yielding a minimal collection of
columns sufficient to span the full column space of $A$ to accuracy
$\epsilon$. Since the matrix $A$ provides an accurate discretization of
the kernel $K(\tau,\omega)$, the selected columns, which correspond to
selected frequencies $\{\omega_l\}_{l=1}^r$, define an accurate
approximate basis $\{K(\tau,\omega_l)\}_{l=1}^r$ of the column space of the Lehmann
representation integral operator. The number $r$ of DLR basis functions
is shown as a function of $\Lambda$ for a few choices of $\epsilon$ in
Figure \ref{fig:dlrrank}, and an example of the selected DLR frequencies
for a given choice of $\Lambda$ and $\epsilon$
is shown in the first panel of Figure \ref{fig:nodes}. The number of
basis functions, called the DLR rank, is observed to scale as $r =
\OO{\log \paren{\Lambda} \log \paren{1/\epsilon}}$.

The DLR coefficients $\wh{g}_l$ of a given Green's function $G$ can be
recovered directly from samples $\{G(\tau_k^s)\}_{k=1}^n$, for some
sampling nodes $\tau_k^s$. This can be done by
solving a linear system
\begin{equation} \label{eq:gsystem}
  \sum_{l=1}^r K(\tau_k^s,\omega_l) \wh{g}_l = G(\tau_k^s),
\end{equation}
for $k = 1,\ldots,n$.
We consider two possible scenarios: (i) Applying the rank revealing row pivoted QR
algorithm to the matrix $K(\tau_i^f,\omega_l)$ yields a set of interpolation nodes
$\{\tau_k\}_{k=1}^r$, called the DLR imaginary time nodes, such that the
coefficients can be recovered from the samples $\{G(\tau_k)\}_{k=1}^r$.
Thus, if $G(\tau)$ can be evaluated at
arbitrary imaginary time points, then we can take $n = r$ and
$\tau_k^s = \tau_k$ in \eqref{eq:gsystem} and solve the resulting small square
linear system to obtain an interpolant of $G(\tau)$. An example of the DLR imaginary time nodes is shown in
the second panel of Figure \ref{fig:nodes}. (ii) If samples of $G$ are given at a collection of $n>r$
scattered or uniform grid points $\tau_k^s$, then \eqref{eq:gsystem} is
an overdetermined system and can be solved by ordinary least
squares fitting.

Since the DLR basis functions are given explicitly, a DLR expansion can
be analytically transformed to the Matsubara
frequency domain. We have
\begin{equation} \label{eq:kmf}
  K(i \nu_n,\omega) = \int_0^\beta K(\tau,\omega) e^{-i \nu_n \tau} \, d\tau
    = \paren{\omega+i\nu_n}^{-1},
\end{equation}
with the Matsubara frequencies given by
\begin{equation} \label{eq:matfreq}
i \nu_n = 
  \begin{cases}
    i (2n+1) \pi / \beta & \text{ for fermionic Green's functions} \\
    i (2n) \pi / \beta & \text{ for bosonic Green's functions}
  \end{cases}
\end{equation}
in unscaled coordinates. In scaled coordinates, we simply set $i \nu_n
\gets i \beta \nu_n$. The DLR expansion in the Matsubara frequency
domain is then
\[\gdlr(i\nu_n) = \sum_{l=1}^r K(i\nu_n,\omega_l) \wh{g}_l.\]
As in the imaginary time domain, there is a set of Matsubara frequencies $\{i
\nu_{n_k}\}_{k=1}^r$, the DLR Matsubara frequency nodes, such that
the coefficients can be recovered from samples $\{G(i
\nu_{n_k})\}_{k=1}^r$ by interpolation; or, one can recover the expansion by least
squares fitting. An example of the DLR
Matsubara frequency nodes for a given choice of $\Lambda$ and $\epsilon$ is shown in the third panel of Figure \ref{fig:nodes}.

For further details on the algorithm used
to obtain the DLR frequencies, imaginary time nodes, and Matsubara
frequencies, as well as a detailed analysis of the accuracy and stability
of the method, we again refer the reader to \cite{kaye21}.

\section{Features and layout of the library} \label{sec:features}

We list the main capabilities of \texttt{libdlr}:
\begin{itemize}
  \item Given a choice of $\Lambda$ and $\epsilon$, obtain the DLR frequencies
    $\{\omega_l\}_{l=1}^r$ characterizing the DLR basis
  \item Obtain the DLR imaginary time and Matsubara frequency grids
  \item Recover the DLR coefficients of an imaginary time Green's
    function from its samples on the DLR grids or from noisy data
  \item Evaluate a DLR expansion at arbitrary points in the imaginary
    time and Matsubara frequency domains
  \item Compute the convolution of imaginary time Green's functions
  \item Solve the Dyson equation in imaginary time or Matsubara
    frequency
\end{itemize}
The computational cost of building the DLR basis and grids is typically
negligible even for large values of $\Lambda$, and subsequent operations
involve standard numerical linear algebra procedures with small matrices.

\texttt{libdlr} is implemented in Fortran,
provides a C header interface,
and includes a Python module,
\texttt{pydlr}. More information on \texttt{pydlr} is contained in
Appendix \ref{app:python}.
The documentation for the library \cite{libdlrdoc}
includes installation intructions, API information, and several
examples. The source code is available as a Git repository
\cite{libdlr}. The subfolder \texttt{libdlr/test} contains example
programs which are run after compilation as tests, and the subfolder
\texttt{libdlr/demo} contains additional example programs. 
An example C program using the C header interface may be found in
\texttt{libdlr/test/ha\_it.c}.

We have made the Julia package \texttt{Lehmann.jl},
which implements functionality similar to \texttt{libdlr}, available in
a separate Git repository \cite{lehmann}. More information on
\texttt{Lehmann.jl} is contained in Appendix \ref{app:julia}, and in the
documentation \cite{lehmanndoc}.

\section{Examples of usage} \label{sec:examples}

We describe two examples illustrating the functionality of
\texttt{libdlr}.
First, we demonstrate recovery of the DLR
coefficients of a Green's function both from its values on the DLR
imaginary time grid and from noisy data on a uniform grid. Second, we demonstrate the process of solving the Dyson equation
self-consistently for the Sachdev-Ye-Kitaev model
\cite{sachdev93,gu20,chowdhury21}. All examples are
implemented in Fortran; Appendices \ref{app:python} and \ref{app:julia} contain analogous
examples implemented using \texttt{pydlr} and \texttt{Lehmann.jl},
respectively.

\subsection{Obtaining a DLR expansion from Green's function samples}

We consider the Green's function given by the Lehmann representation
\eqref{eq:lehmann}
with spectral density
\begin{equation} \label{eq:rhosc}
  \rho(\omega) = \frac{2}{\pi} \sqrt{1-\omega^2} \, \theta(1-\omega^2).
\end{equation}
Here, $\theta$ is the Heaviside function.
In practical calculations,
the spectral density is usually not known, so
one must recover the DLR from samples of the Green's function itself.
We use a Green's function with known spectral density for illustrative purposes,
since we can evaluate it with high accuracy
by numerical integration of \eqref{eq:lehmann} and thereby test our
results.

Since $\supp \rho \subset [-1,1]$, we can take the frequency support cutoff
$\omega_{\max} = 1$. Then $\Lambda = \beta \omega_{\max} \geq \beta$ is
a sufficient high energy cutoff. In practice, $\omega_{\max}$ can
typically be estimated on physical grounds, giving an estimate of
$\Lambda$. Calculations can then be converged with respect to
$\Lambda$ to ensure accuracy.
The error tolerance $\epsilon$ should be chosen based on the desired
accuracy, in order to obtain the smallest possible number of basis
functions. In this example, we take $\beta = 1000$, and fix $\Lambda =
1000$.

We note also that \texttt{libdlr} routines work by default
with matrix-valued Green's functions $G_{ij}$, where $i$ and $j$ are
typically orbital indices. However, in all examples
presented here, we work with scalar-valued Green's functions, and
therefore set the parameter \texttt{n} determining the number of orbital
indices to $1$.

\subsubsection{Recovery of DLR from imaginary time grid values}

We first consider recovery of the DLR coefficients $\wh{g}_l$ from
samples of $G(\tau)$ on the DLR imaginary time nodes $\tau_k$. This is
the first scenario mentioned in Section \ref{sec:dlr}, and
generates an interpolant at the nodes $\tau_k$. Figure
\ref{fig:codescit} shows a condensed version of a Fortran code
implementing the example using \texttt{libdlr}. In this and all other
sample Fortran codes, 
we do not show variable allocations or steps which do not involve
\texttt{libdlr} subroutines, as indicated in the code comments. A
complete Fortran code demonstrating this example can be found in the
file \texttt{libdlr/demo/sc\_it.f90}. The file
\texttt{libdlr/demo/sc\_mf.f90} contains a demonstration of the
process of obtaining a DLR from values of a Green's function on the DLR
Matsubara frequency nodes.

\begin{figure}
\begin{minted}{fortran}
      ! Set parameters and build DLR basis, imaginary time grid

      lambda = 1000.0d0 ! DLR high energy cutoff
      eps = 1.0d-10     ! DLR error tolerance
      xi = -1           ! Fermionic Green's function
      n = 1             ! # orbital indices

      call dlr_it_build(lambda,eps,r,dlrrf,dlrit)

      ! Recovery of DLR from DLR imaginary time nodes (assume values of G
      ! on DLR imaginary time grid dlrit stored in array g)

      call dlr_it2cf_init(r,dlrrf,dlrit,it2cf,it2cfp)
      call dlr_it2cf(r,n,it2cf,it2cfp,g,gc)

      ! Evaluate on imaginary time output grid of nout points

      call eqpts_rel(nout,tout) ! Equispaced grid in relative format

      do i=1,nout
        call dlr_it_eval(r,n,dlrrf,gc,tout(i),gout_it(i))
      enddo

      ! Evaluate on Matsubara frequency output grid with cutoff index
      ! nmfmax

      do nmf=-nmfmax,nmfmax
        call dlr_mf_eval(r,n,dlrrf,xi,gc,nmf,gout_mf(nmf))
      enddo
\end{minted}
    \caption{\texttt{libdlr} Fortran code to obtain DLR from values of a
    Green's function on the
    DLR imaginary time grid.  To emphasize the usage of \texttt{libdlr}
    subroutines, we do not show variable allocations or the external
    procedures used to sample the Green's function, as indicated in the
    code comments.}
    \label{fig:codescit}
\end{figure}

We first set $\Lambda$ and $\epsilon$, and then build the DLR basis by
obtaining the DLR real frequencies $\omega_l$, stored in the array
\texttt{dlrrf}, using the subroutine \texttt{dlr\_it\_build}. This
subroutine also produces the $r$ DLR imaginary time nodes $\tau_k$,
which are stored in the array \texttt{dlrit}.

Next, we assume that the values $G(\tau_k)$ have been obtained by some
external procedure, and stored in the array \texttt{g}. In this case,
we used numerical integration with the known spectral function to obtain
the samples. To obtain the DLR coefficients $\wh{g}_l$, one must solve the linear
system \eqref{eq:gsystem}, with $n = r$. The subroutine
\texttt{dlr\_it2cf\_init} initializes this procedure by computing the LU
factorization of the system matrix. The LU factors and pivots are stored
in the arrays \texttt{it2cf} and \texttt{it2cfp}, respectively. The linear solve is then carried out
by the subroutine \texttt{dlr\_it2cf}, which returns the DLR
coefficients $\wh{g}_l$ in the array \texttt{gc}.

We can then evaluate the DLR expansion on output grids in imaginary time
and Matsubara frequency. The subroutine \texttt{eqpts\_rel} generates a
uniform grid of imaginary time points in the relative format employed by
the library (see Appendix \ref{app:rel} for an explanation of the
relative format) and \texttt{dlr\_it\_eval} evaluates the expansion. The
subroutine \texttt{dlr\_mf\_eval} evaluates the DLR expansion in the
Matsubara frequency domain.

\subsubsection{Recovery of DLR from noisy data}

We next consider DLR fitting from noisy data, the second scenario
mentioned in Section \ref{sec:dlr}. A condensed sample code is given in
Figure \ref{fig:codescfit}, and a complete example can be found in
the file \texttt{libdlr/demo/sc\_it\_fit.f90}. We assume noisy
samples $G(\tau_k^s)$ have been obtained by an external procedure and
stored in the array $\texttt{g}$. The subroutine \texttt{dlr\_it\_fit}
fits a DLR expansion to the data by solving the overdetermined system
\eqref{eq:gsystem}.  The resulting DLR coefficients, stored in the array
\texttt{gc}, can be used to evaluate the DLR expansion as in the
previous example.

\begin{figure}
\begin{minted}{fortran}
      ! Set parameters and build DLR basis, imaginary time grid

      lambda = 1000.0d0 ! DLR high energy cutoff
      eps = 1.0d-4      ! DLR error tolerance
      n = 1             ! # orbital indices

      call dlr_it_build(lambda,eps,r,dlrrf,dlrit)

      ! Recovery of DLR from noisy data on uniform grid (assume data
      ! given at m equispaced grid points t stored in array g)

      call dlr_it_fit(r,n,dlrrf,m,t,g,gc)
\end{minted}
    \caption{\texttt{libdlr} Fortran code to obtain DLR from noisy data
    on a uniform grid.}
    \label{fig:codescfit}
\end{figure}

\subsubsection{Numerical results}

\begin{figure}
  \centering
  \includegraphics[width=0.32\textwidth]{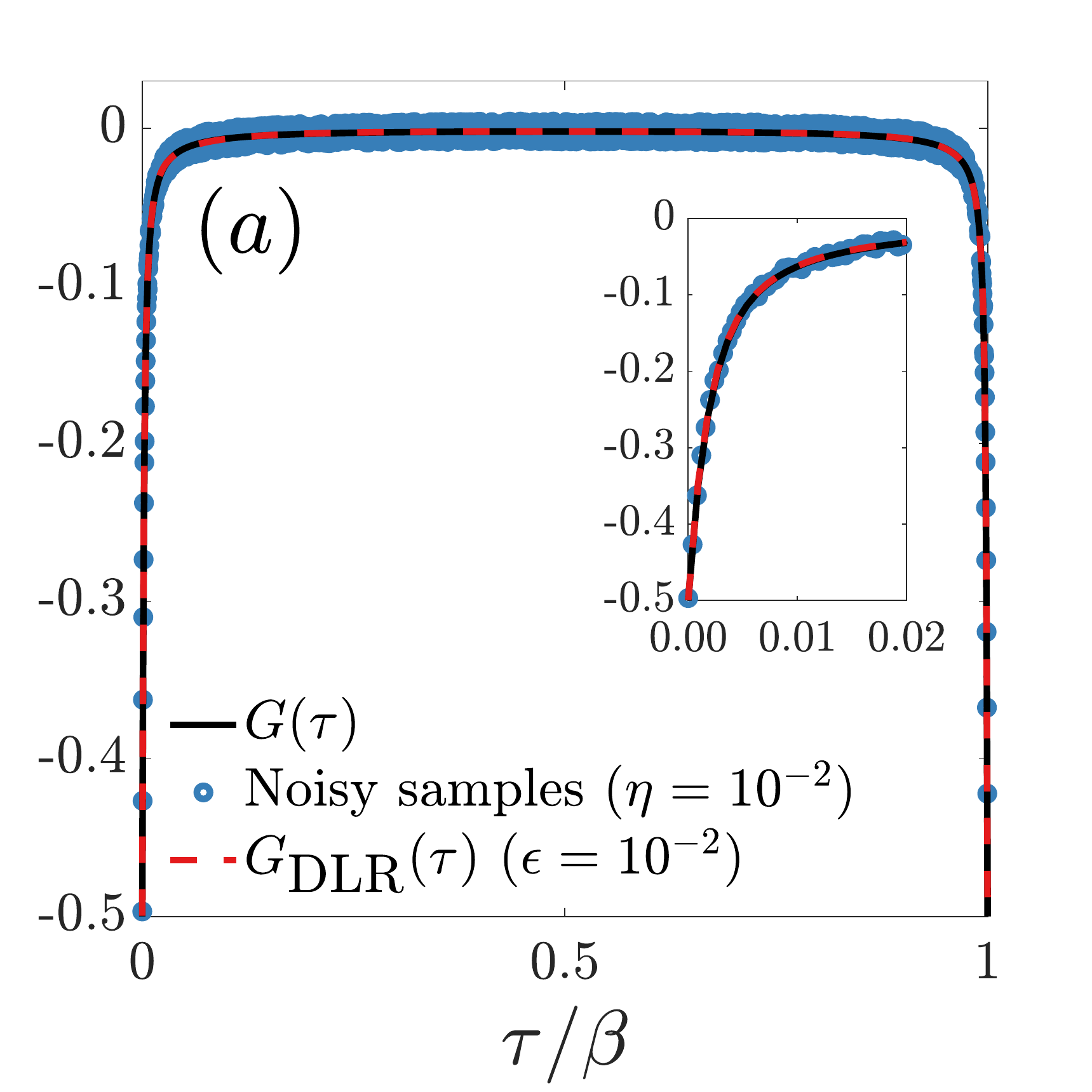}
  \includegraphics[width=0.32\textwidth]{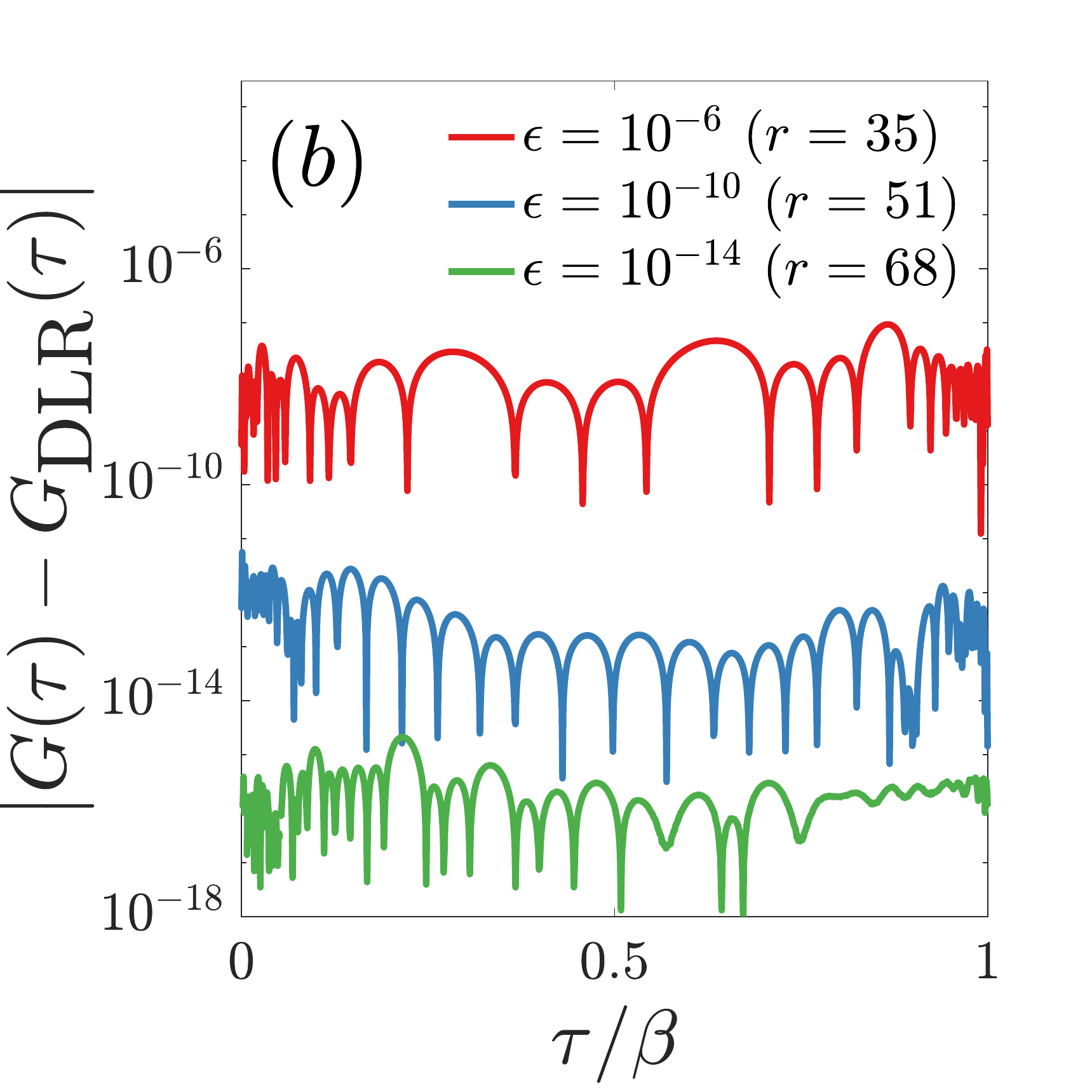}
  \includegraphics[width=0.32\textwidth]{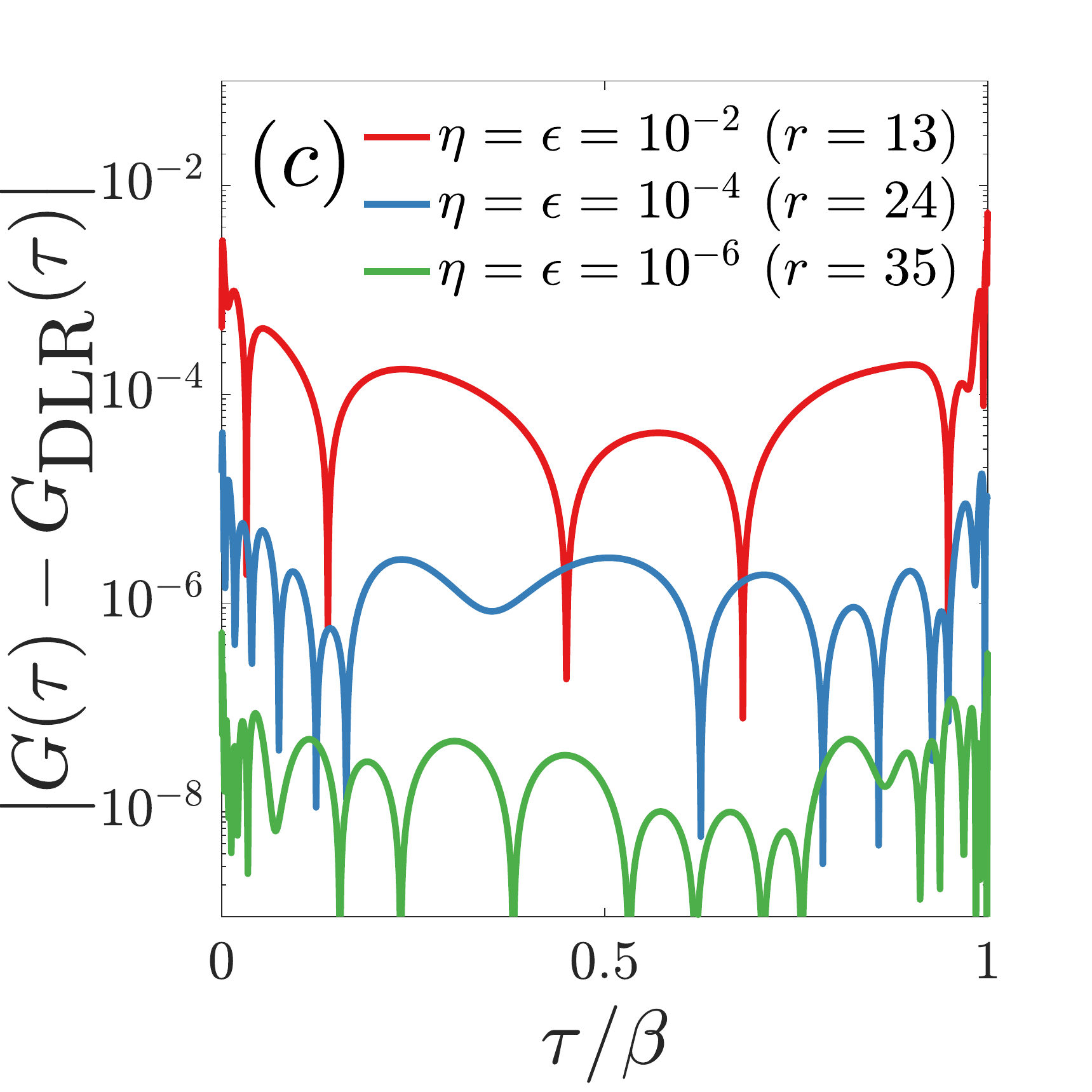}
  \caption{DLR expansion of a Green's function with 
  spectral density \eqref{eq:rhosc}, for $\beta = 1000$, $\Lambda = 1000$. (a) Noisy imaginary time data (noise of
  magnitude $\eta = 10^{-2}$) with DLR fit ($\epsilon = 10^{-2}$, $r =
  13$). (b) Error of
  DLR expansion obtained from imaginary time grid sampling, with
  different choices of $\epsilon$. DLR rank $r$ obtained for each choice
  of $\epsilon$ also indicated. (c) Error of DLR expansion obtained from
  fit of $n = 2500$ noisy uniform grid samples, with different noise
  levels $\eta$, and $\epsilon = \eta$.
  }
  \label{fig:sc}
\end{figure}

Figure \ref{fig:sc} presents some numerical results for the two
examples. In Figure \ref{fig:sc}a, we plot the Green's function
$G(\tau)$ with spectral density \eqref{eq:rhosc} and
$\beta = 1000$, along with noisy data obtained on a uniform grid of $n =
2500$ points by adding uniform random numbers of magnitude $\eta =
10^{-2}$ to accurate values computed by numerical integration. We see
that the DLR fit $\gdlr$ to this data with $\epsilon = 10^{-2}$ agrees
well with $G(\tau)$.  Pointwise errors for this example, and
examples with different noise levels $\eta$, are given in Figure
\ref{fig:sc}c. In all cases we take $\epsilon = \eta$, and observe that
the DLR fitting process is stable; the fitting process does not
introduce a significant error above the magnitude of the noise.
Pointwise errors for the DLR obtained from numerically exact samples on
the DLR imaginary time grid are shown in Figure \ref{fig:sc}b, and we see
that the error is well controlled by $\epsilon$.

\subsection{Solving the Sachdev-Ye-Kitaev model}

The Dyson equation for a given self-energy $\Sigma$ can be written in
the Matsubara frequency domain as
\begin{equation} \label{eq:dysonmf}
   G^{-1}(i\nu_n) = G^{-1}_0(i\nu_n) - \Sigma(i\nu_n). 
\end{equation}
Here $G_0(i \nu_n) = (i\nu_n+\mu)^{-1}$ is the free particle
Green's function, with chemical potential $\mu$. This equation can be solved for $G$ by pointwise inversion on
the $r$ DLR Matsubara frequency nodes.

In typical applications, $\Sigma$ is a function of $G$, $\Sigma =
\Sigma\brak{G}$, which
is most easily evaluated in the imaginary time domain. In this case,
\eqref{eq:dysonmf} becomes nonlinear, and must be
solved by self-consistent iteration. The standard method is to solve the Dyson
equation using \eqref{eq:dysonmf} for a given iterate of $\Sigma$, transform the solution $G$
to the imaginary time domain to evaluate a new iterate of $\Sigma$, and then transform
$\Sigma$ back to the Matsubara frequency domain to solve
\eqref{eq:dysonmf} for the next iteration. The standard implementation
of this approach 
represents $G$ and $\Sigma$ on fine equispaced grids in
imaginary time, and uses the fast Fourier transform to move between the
imaginary time and Matsubara frequency domains. The
process can be carried out significantly more efficiently using the DLR.

As an example, we consider the Dyson equation given by
\eqref{eq:dysonmf} with the Sachdev-Ye-Kitaev (SYK) \cite{sachdev93,gu20,chowdhury21} self-energy
\begin{equation}\label{eq:sigmasyk}
  \Sigma(\tau) = J^2 G^2(\tau) G(\beta-\tau).
\end{equation}
Here $J$ is a coupling constant.
We consider this example with $\beta
= 1000$, $\mu = 0$, and $J=1$, and set $\Lambda = 5000$. We note that $G$ is a fermionic
Green's function, so we use a fermionic Matsubara frequency grid.

Figure \ref{fig:codedyson} gives sample code for a
solver which uses a weighted
fixed point iteration to handle the nonlinearity:
\begin{equation}\label{eq:mixing}
\Sigma^{(m+1)} = \Sigma[w \, G^{(m)} + (1-w) \, G^{(m-1)}].
\end{equation}
Here, $m$ refers to the iterate, and $w$ is a weighting parameter which
can be selected to improve convergence; we use $w = 0.3$. A complete code
demonstrating this example can be
found in the file \texttt{libdlr/demo/syk\_mf.f90}.

\begin{figure}
\begin{minted}{fortran}
      ! Set DLR parameters and build DLR basis, imaginary time grid,
      ! Matsubara frequency grid

      lambda = 5000.0d0   ! DLR high energy cutoff
      eps = 1.0d-10       ! DLR error tolerance
      nmfmax = 5000       ! DLR Matsubara frequency cutoff
      xi = -1             ! Fermionic Green's function
      n = 1               ! # orbital indices

      call dlr_it_build(lambda,eps,r,dlrrf,dlrit)
      call dlr_mf(nmfmax,r,dlrrf,xi,dlrmf)

      ! Set problem parameters and parameters for nonlinear iteration

      beta = 1000.0d0   ! Inverse temperature
      mu = 0.0d0        ! Chemical potential
      j = 1.0d0         ! Coupling constant

      numit = 100       ! Max # weighted fixed point iterations
      fptol = 1.0d-10   ! Fixed point tolerance
      w = 0.3d0         ! Fixed point weighting parameter

      ! Initialize transformations between imaginary time, Matsubara
      ! frequency grids and DLR coefficients
      
      call dlr_it2cf_init(r,dlrrf,dlrit,it2cf,it2cfp)
      call dlr_cf2it_init(r,dlrrf,dlrit,cf2it)

      call dlr_mf2cf_init(nmfmax,r,dlrrf,dlrmf,xi,mf2cf,mf2cfp)
      call dlr_cf2mf_init(r,dlrrf,dlrmf,xi,cf2mf)

      call dlr_it2itr_init(r,dlrrf,dlrit,it2cf,it2cfp,it2itr)

      ! Get free particle Green's function on Matsubara frequency grid

      call g0_mf(beta,r,dlrmf,mu,g0)

      ! Get free particle Green's function on imaginary time grid for
      ! initial guess in nonlinear iteration
      
      call g0_it(beta,r,dlrit,mu,g)

      ! Weighted fixed point iteration

      do i=1,numit

        gprev = g                                 ! Store previous iterate
        call sigfun(r,j,it2itr,g,sig)             ! Evaluate self-energy

        call dlr_it2cf(r,n,it2cf,it2cfp,sig,sigc) ! Imaginary time grid -> DLR coeffs
        call dlr_cf2mf(r,n,cf2mf,sigc,sigmf)      ! DLR coeffs -> Matsubara frequency grid

        call dyson_mf(beta,r,n,g0,sigmf,gmf)      ! Solve Dyson equation

        call dlr_mf2cf(r,n,mf2cf,mf2cfp,gmf,gc)   ! Matsubara frequency grid -> DLR coeffs
        call dlr_cf2it(r,n,cf2it,gc,g)            ! DLR coeffs -> imaginary time grid

        if (maxval(abs(gprev-g))<fptol) then      ! Check self-consistency 
          return
        else
          g = w*g + (1.0d0-w)*gprev               ! Reweight
        endif
      enddo
\end{minted}
    \caption{\texttt{libdlr} Fortran code to solve the SYK model.}
    \label{fig:codedyson}
\end{figure}

As in the previous example, the code begins by obtaining the DLR
frequencies and imaginary time grid. Here, the DLR
Matsubara frequency grid is also built, using the subroutine
\texttt{dlr\_mf}, with a Matsubara frequency cutoff
\texttt{nmax}. This cutoff should
in principle be taken larger and larger until the selected
DLR Matsubara frequencies no longer change, but in practice we find that
setting it to $\Lambda$ is usually sufficient (see the discussion in
\cite[Sec. IV D]{kaye21}).

The next several lines define physical problem parameters, and
parameters for the weighted fixed point iteration. Then, several initialization
routines are called which prepare transformations between the imaginary time and
Matsubara frequency grid representations of the Green's function, and
the DLR coefficients. These transformations are used to convert between
imaginary time and Matsubara frequency representations in the
self-consistent iteration. The subroutine \texttt{dlr\_it2itr\_init}
prepares a transformation between a
Green's function $G(\tau)$ on the imaginary time grid and its reflection
$G(\beta-\tau)$ on the same grid, which is needed to evaluate the SYK
self-energy.

The free particle Green's function $G_0$ appearing in \eqref{eq:dysonmf}
is then evaluated on the Matsubara frequency grid, since it appears in
the Dyson equation, and on the imaginary time grid, to serve as an initial guess in the iteration.
In the weighted fixed point iteration, the self-energy is evaluated on
the imaginary time grid, and then transformed to the Matsubara frequency
domain, where the Dyson equation is solved. Then the result is
transformed back to the imaginary time grid, where the self-consistency
of the solver is checked. Figure \ref{fig:sigfun} shows the subroutine used to evaluate the SYK
self-energy in imaginary time.
Once self-consistency is reached, the solution
is returned on the imaginary time grid. It can be expanded in a DLR
and evaluated in imaginary time or Matsubara frequency as in the
previous example.

\begin{figure}
\begin{minted}{fortran}
      subroutine sigfun(r,j,it2itr,g,sig)

      ! Evaluate SYK self-energy

      n = 1                             ! Scalar-valued G
      call dlr_it2itr(r,n,it2itr,g,gr)  ! Get G(beta-tau)
      sig = j*j*g*g*gr                  ! Get Sigma(tau)

      end subroutine sigfun
\end{minted}
    \caption{Evaluation subroutine for SYK self-energy, used in the SYK
    solver shown in Figure \ref{fig:codedyson}.}
    \label{fig:sigfun}
\end{figure}

\begin{figure}
  \centering
  \includegraphics[width=0.4\textwidth]{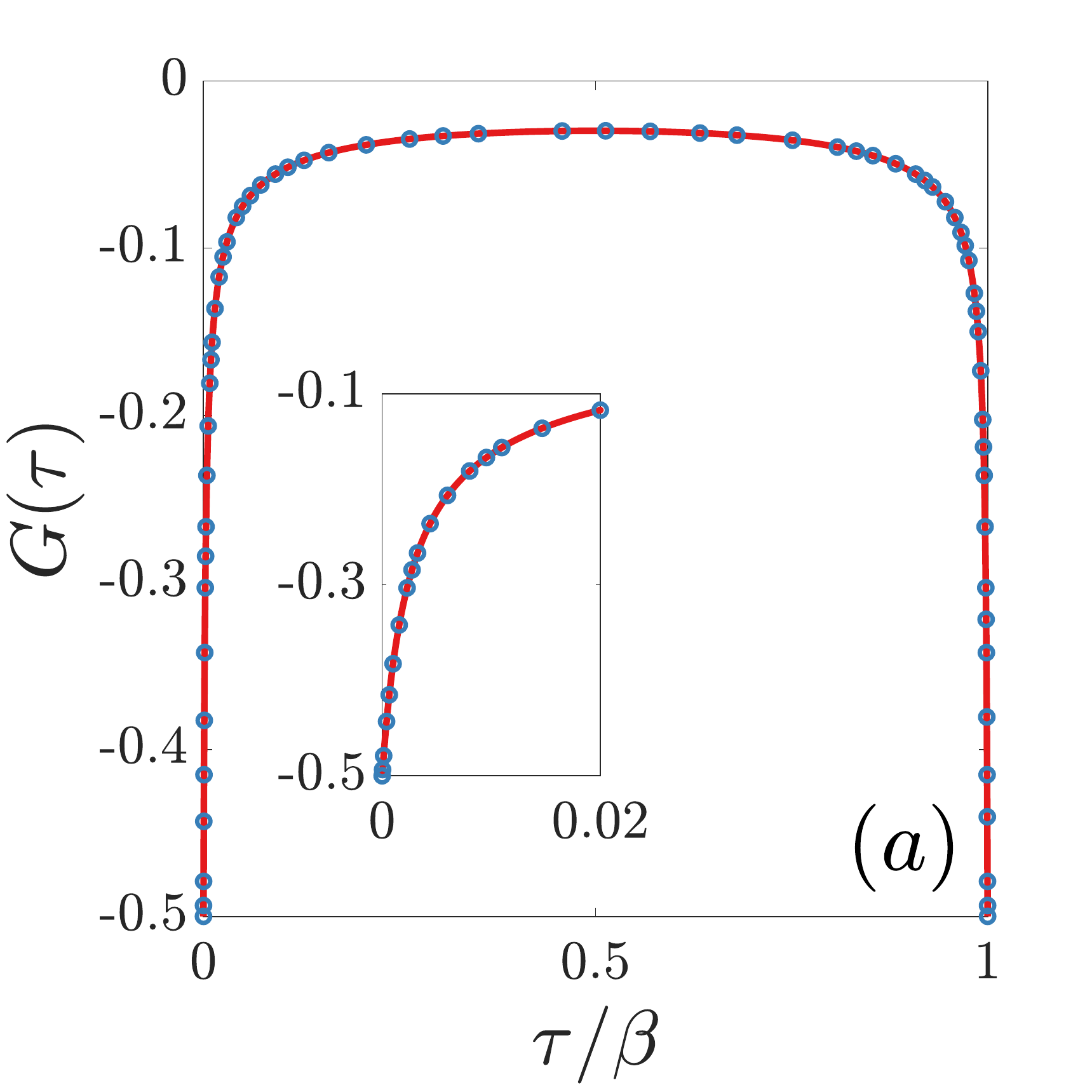}
  \hspace{5ex}
  \includegraphics[width=0.4\textwidth]{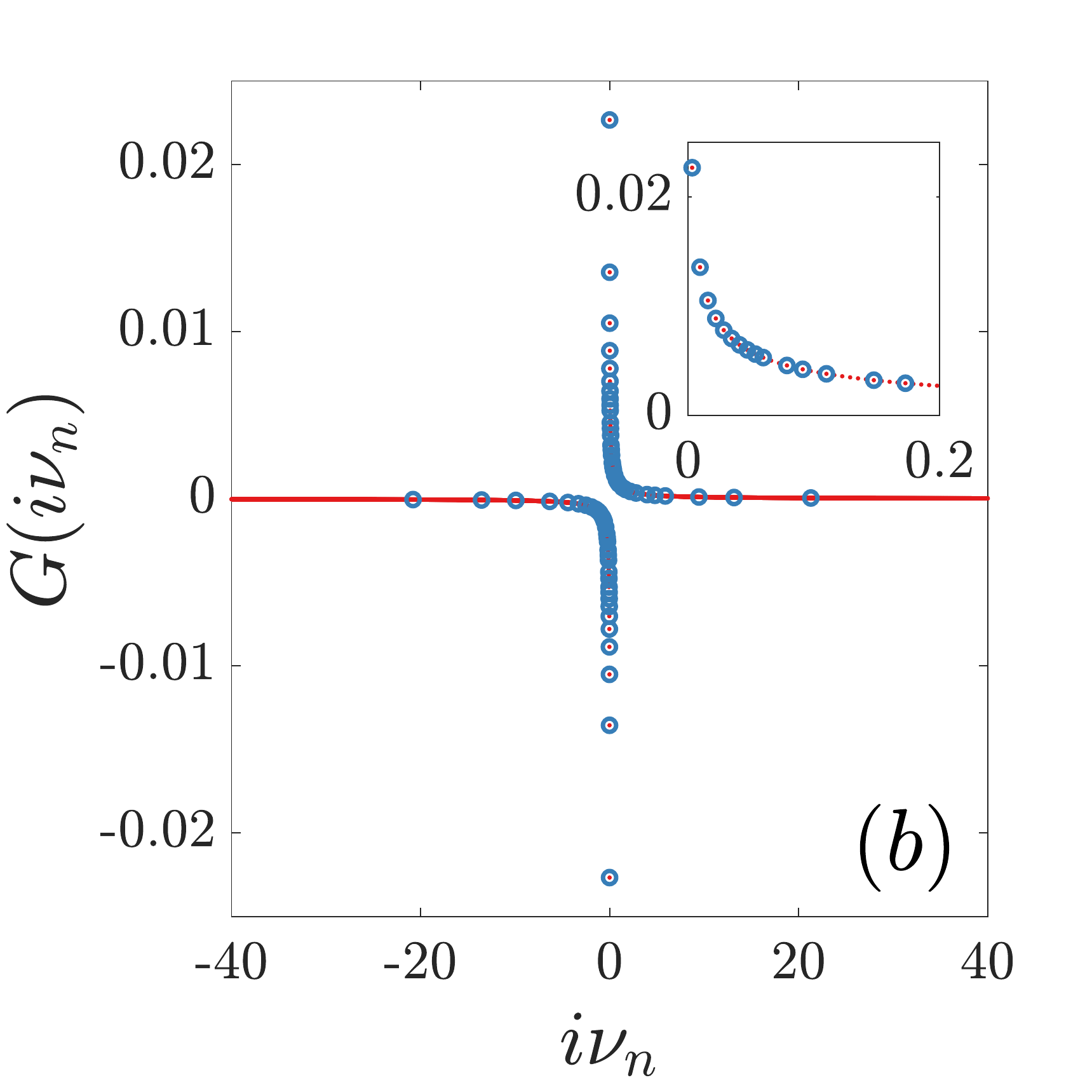}
  \caption{Solution of the SYK model with $\beta =
  1000$, $\mu = 0$, and $J=1$. (a) $G(\tau)$, with $r = 66$ DLR imaginary time nodes for
  $\epsilon = 10^{-10}$ accuracy indicated by the blue circles. (b) $G(i
  \nu_n)$, with DLR Matsubara frequency nodes.}
  \label{fig:syk}
\end{figure}

Plots of the solution and of the DLR nodes are given in Figure \ref{fig:syk}, both in
imaginary time and Matsubara frequency. For $\Lambda = 5000$ and $\epsilon = 10^{-10}$, there
are $r = 66$ DLR nodes discretizing each domain.

We remark that the Dyson equation \eqref{eq:dysonmf} may be written in the imaginary
time domain as
\begin{equation} \label{eq:dysonit}
    G(\tau) - \int_0^\beta d\tau' G_0(\tau-\tau') \int_0^\beta d\tau'' 
    \Sigma(\tau'-\tau'') G(\tau'') = G_0(\tau).
\end{equation}
In \cite{kaye21}, an efficient method is presented of solving
\eqref{eq:dysonit} by discretizing the
convolutions using the DLR. This approach may be advantageous in certain
cases, and can also be implemented using \texttt{libdlr}. A
demonstration for the SYK model is given in the file
\texttt{libdlr/demo/syk\_it.f90}.

We also note that the convergence properties of nonlinear iteration in
the self-consistent solution of the Dyson equation are problem-dependent.
For the SYK model (\ref{eq:sigmasyk}), which belongs to a particular subclass of approximations \cite{https://doi.org/10.48550/arxiv.2203.11083},
we find that the weighted fix point iteration (\ref{eq:mixing}) may fail to converge or converge to spurious local fixed points, particularly at extremely low temperatures.
Many standard strategies can be used to improve convergence,
including the careful selection of an initial guess by
adiabatic variation of parameters like $\beta$ or $\mu$, mixing, and more
sophisticated nonlinear iteration procedures. Two other methods
have been observed to improve performance in certain difficult cases: the use of explicitly
symmetrized DLR grids, and mild oversampling in the Matsubara frequency
domain followed by least squares fitting to a DLR expansion. These are
topics of our current research, and will be reported on in detail at a later date.
Lastly, we have observed that solving \eqref{eq:dysonit} directly in the
imaginary time domain, as mentioned in the previous paragraph, tends to
lead to more robust convergence.

\section{Conclusion} \label{sec:conclusion}

\texttt{libdlr} facilitates the efficient representation and
manipulation of imaginary time Green's functions using the DLR.
In this framework, working with imaginary time Green's functions is as
simple as with standard discretizations, but involves many fewer degrees
of freedom, and user-tuneable accuracy down to the level of machine
precision.

As a result, we anticipate the use of the DLR as a basic working tool a
variety of equilibrium applications, including continous-time quantum
Monte Carlo \cite{Gull:2011lr}, dynamical mean-field theory
\cite{Georges:1996aa}, self-consistent perturbation theory in both
quantum chemistry (GF2) \cite{Dahlen:2005aa, Phillips:2014aa,
PhysRevB.97.115164, doi:10.1063/5.0003145} and condensed matter physics
\cite{gf2:periodic}, as well as Hedin's GW approximation
\cite{Hedin:1965aa, 10.3389/fchem.2019.00377}, including vertex
corrections \cite{PhysRevB.104.085109, PhysRevB.90.115134}. In
\cite{kaye21_2},
the DLR was used to discretize imaginary time variables in equilibrium
real time contour Green's functions. In nonequilibrium calculations involving
two-time Green's functions, it can replace the equispaced imaginary time grids
currently in use \cite{Stan:2009ab, Schuler:2020uy}, and further improve
the efficiency of algorithms making use of low rank compression to
reduce the cost of time propagation \cite{10.21468/SciPostPhys.10.4.091}.

\section*{Acknowledgements}

H.U.R.S.\ acknowledges financial support from the ERC synergy grant (854843-FASTCORR).
The Flatiron Institute is a division of the Simons Foundation.

\appendix

\section{Python module \texttt{pydlr}} \label{app:python}

The library \texttt{libdlr} provides a stand-alone Python module
\texttt{pydlr}, so that small-scale tests can be easily carried out in
Python.
The \texttt{libdlr} documentation \cite{libdlrdoc} includes usage examples and API
instructions for \texttt{pydlr}.
Here, we show how the examples discussed in Section
\ref{sec:examples} can be implemented using \texttt{pydlr}.
\begin{enumerate}
  \item
The first example, demonstrating recovery of the DLR coefficients
$\wh{g}_l$ from values of a Green's function at the DLR imaginary time
nodes $\tau_k$, is shown in Figure \ref{fig:pythoncodeeval}. We also
demonstrate the evaluation of the DLR at arbitrary imaginary time points. We again use the example of the Green's function with
    spectral density \eqref{eq:rhosc}.

\item
The second example, demonstrating recovery of the DLR coefficients from
noisy data in imaginary time, is shown in Figure \ref{fig:pythoncodefit}.

\item
The final example, demonstrating the iterative solution of the SYK
    model, is shown in Figure \ref{fig:pythoncodedyson}.
\end{enumerate}

\begin{figure}
\begin{minted}{python}
import numpy as np
from pydlr import dlr
from pydlr.utils import analytic_bethe_G_tau as true_G_tau

beta = 1000.                                # Inverse temperature
d = dlr(lamb=1000., eps=1e-14)              # Initialize DLR object
tau_k = d.get_tau(beta)                     # DLR imaginary time points

G_k = true_G_tau(tau_k, beta)               # Evaluate known G at tau_k
G_x = d.dlr_from_tau(G_k)                   # DLR coeffs from G_k

tau_i = np.linspace(0, beta, num=40)        # Equispaced tau grid
G_i = d.eval_dlr_tau(G_x, tau_i, beta)      # Evaluate DLR at tau_i
\end{minted}
\caption{\texttt{pydlr} Python code to obtain DLR from values of the
  Green's function with spectral density \eqref{eq:rhosc} on the DLR
  imaginary time grid, and to evaluate the DLR at arbitrary points in imaginary time.}
    \label{fig:pythoncodeeval}
\end{figure}

\begin{figure}
\begin{minted}{python}
import numpy as np
from pydlr import dlr

eta = 1e-3                                 # Noise level
beta = 1000                                # Inverse temperature
d = dlr(lamb=1000., eps=1e-14)             # Initialize DLR object

# Generate noisy data

from pydlr.utils import analytic_bethe_G_tau as true_G_tau

tau_i = np.linspace(0, beta, num=100)
G_i = true_G_tau(tau_i, beta)
G_i_noisy = G_i + eta*(np.random.random(G_i.shape)-.5)

# Fit DLR coeffs from noisy data

G_x = d.lstsq_dlr_from_tau(tau_i, G_i_noisy, beta)

# Evaluate DLR

G_i_fit = d.eval_dlr_tau(G_x, tau_i, beta)
\end{minted}
\caption{\texttt{pydlr} Python code to obtain DLR from noisy data on a
  uniform grid.}
    \label{fig:pythoncodefit}
\end{figure}

\begin{figure}
\begin{minted}{python}
import numpy as np
from pydlr import dlr

def syk_sigma_dlr(d, G_x, beta, J=1.):

    tau_k = d.get_tau(beta)                         # DLR imaginary time nodes
    tau_k_rev = beta - tau_k                        # Reversed imaginary time nodes
    
    G_k = d.tau_from_dlr(G_x)                       # G at tau_k
    G_k_rev = d.eval_dlr_tau(G_x, tau_k_rev, beta)  # G at beta - tau_k

    Sigma_k = J**2 * G_k**2 * G_k_rev               # SYK self-energy in imaginary time
    Sigma_x = d.dlr_from_tau(Sigma_k)               # DLR coeffs of self-energy
    
    return Sigma_x

def solve_syk_with_fixpoint_iter(d, mu, beta, tol=1e-14, mix=0.3, maxiter=100):
    
    Sigma_q = np.zeros((len(d), 1, 1))                            # Initial guess

    for iter in range(maxiter):

        G_q = d.dyson_matsubara(np.array([[-mu]]), Sigma_q, beta) # Solve Dyson

        G_x = d.dlr_from_matsubara(G_q, beta)                     # Get DLR coeffs
        Sigma_x_new = syk_sigma_dlr(d, G_x, beta)                 # Compute self-energy
        Sigma_q_new = d.matsubara_from_dlr(Sigma_x_new, beta)     # Transform self-energy to Matsubara frequency

        if np.max(np.abs(Sigma_q_new - Sigma_q)) < tol: break     # Check self-consistency
        Sigma_q = mix * Sigma_q_new + (1-mix) * Sigma_q           # Linear mixing
        
    return G_q

d = dlr(lamb=5000., eps=1e-14)
G_q = solve_syk_with_fixpoint_iter(d, mu=0., beta=1000.)
\end{minted}
\caption{\texttt{pydlr} Python code to solve the SYK model.}
    \label{fig:pythoncodedyson}
\end{figure}

\section{Julia module \texttt{Lehmann.jl}} \label{app:julia}

The stand-alone package \texttt{Lehmann.jl} provides a pure Julia
implementation of the DLR. 
The \texttt{Lehmann.jl} documentation \cite{lehmanndoc} includes usage examples and API
instructions. 
The examples discussed in Section
\ref{sec:examples} and Appendix \ref{app:python} can be implemented
using \texttt{Lehmann.jl}, as shown in Figures \ref{fig:juliacodeeval},
\ref{fig:juliacodefit}, and \ref{fig:juliacodedyson}. 

\begin{figure}
\begin{minted}{julia}
using Lehmann

beta = 1000.0                                          # Inverse temperature
Euv = 1.0                                              # We define Lambda = Euv * beta
d = DLRGrid(Euv, beta, rtol = 1e-14, isFermi = true)   # Initialize DLR object (fermionic Green's function)
tau_k = d.tau                                          # DLR imaginary time points

G_k = true_G_tau(tau_i, d.beta)                        # Evaluate known G at tau_k
G_x = tau2dlr(d, G_k)                                  # DLR coeffs from G_k

tau_i = collect(LinRange(0, d.beta, 40))               # Equispaced tau grid
G_i = dlr2tau(d, G_x, tau_i)                           # Evaluate DLR at tau_i
\end{minted}
\caption{\texttt{Lehmann.jl} Julia code to obtain DLR from values of the
  Green's function with spectral density \eqref{eq:rhosc} on the DLR
  imaginary time grid, and to evaluate the DLR at arbitrary points in imaginary time.}
    \label{fig:juliacodeeval}
\end{figure}

\begin{figure}
\begin{minted}{julia}
using Lehmann

eta = 1e-3                                             # Noise level
beta = 1000                                            # Inverse temperature
Euv = 1.0                                              # Lambda = Euv * beta
d = DLRGrid(Euv, beta, rtol = 1e-14, isFermi = true)   # Initialize DLR object

# Generate noisy data

tau_i = collect(LinRange(0, d.beta, 100))
G_i = true_G_tau(tau_i, d.beta)
G_i_noisy = G_i .+ eta .* (rand(length(G_i)) .- 0.5)

# Fit DLR coeffs from noisy data

G_x = tau2dlr(d, G_i_noisy, tau_i)

# Evaluate DLR

G_i_fit = dlr2tau(d, G_x, tau_i)
\end{minted}
\caption{\texttt{Lehmann.jl} Julia code to obtain DLR from noisy data on a
  uniform grid.}
    \label{fig:juliacodefit}
\end{figure}

\begin{figure}
\begin{minted}{julia}
using Lehmann

function syk_sigma_dlr(d, G_x, J = 1.0)

    tau_k = d.tau                             # DLR imaginary time nodes
    tau_k_rev = d.beta .- tau_k               # Reversed imaginary time nodes

    G_k = dlr2tau(d, G_x)                     # G at tau_k
    G_k_rev = dlr2tau(d, G_x, tau_k_rev)      # G at beta - tau_k

    Sigma_k = J .^ 2 .* G_k .^ 2 .* G_k_rev   # SYK self-energy in imaginary time
    Sigma_x = tau2dlr(d, Sigma_k)             # DLR coeffs of self-energy

    return Sigma_x
end

function solve_syk_with_fixpoint_iter(d, mu, tol = d.rtol, mix = 0.3, maxiter = 100)

    Sigma_q = zeros(length(d))                              # Initial guess
    G_q = zeros(ComplexF64, length(d))
    for iter in 1:maxiter

        G_q .= -1 ./ (d.omegaN * 1im .- mu .+ Sigma_q)      # Solve Dyson

        G_x = matfreq2dlr(d, G_q)                           # Get DLR coeffs
        Sigma_x_new = syk_sigma_dlr(d, G_x)                 # Compute self-energy
        Sigma_q_new = dlr2matfreq(d, Sigma_x_new)           # Transform self-energy to Matsubara frequency 

        if maximum(abs.(Sigma_q_new .- Sigma_q)) < tol      # Check self-consistency
            break
        end

        Sigma_q = mix * Sigma_q_new + (1 - mix) * Sigma_q   # Linear mixing

    end
    return G_q
end

d = DLRGrid(Euv = 5.0, beta = 1000.0, isFermi = true, rtol = 1e-14)
G_q = solve_syk_with_fixpoint_iter(d, 0.0)
\end{minted}
\caption{\texttt{Lehmann.jl} Julia code to solve the SYK model.}
    \label{fig:juliacodedyson}
\end{figure}

\section{Stable kernel evaluation and relative format for imaginary time points} \label{app:rel}

In \texttt{libdlr}, we work in the dimensionless variables defined at
the beginning of Section \ref{sec:dlr}, in which the kernel
$K(\tau,\omega)$ is given by
\begin{equation} \label{eq:k1}
  K(\tau,\omega) = \frac{e^{-\tau \omega}}{1 + e^{-\omega}}
\end{equation}
for $\tau \in [0,1]$ and $\omega \in (-\infty,\infty)$.
When $\omega \geq 0$, this formula is numerically stable. If $\omega
\ll 0$, it can overflow, and we instead use the mathematically
equivalent formula
\begin{equation} \label{eq:k2}
  K(\tau,\omega) = \frac{e^{(1-\tau) \omega}}{1 + e^\omega}.
\end{equation}

This formula, however, leads to another problem: when $\abs{\omega}$ is
large, catastrophic
cancellation in the calculation of $1-\tau$ for $\tau$ near $1$ leads to
a loss of accuracy in floating point arithmetic.

To fix this problem, we define a \emph{relative}
format for representing imaginary time values $\tau \in (0.5,1)$. Rather than
representing them directly, which we refer to as the \emph{absolute} format,
we instead store $\tau^* = \tau-1 \in (-0.5,0)$ to full relative accuracy. Then the kernel is evaluated as
$K(-\tau^*,-\omega)$. If $\omega > 0$, \eqref{eq:k2} then yields the
numerically stable formula
\[K(-\tau^*,-\omega) = \frac{e^{-(1+\tau^*) \omega}}{1 + e^{-\omega}}.\]
Since
\[\frac{e^{-(1+\tau^*) \omega}}{1 + e^{-\omega}} =
\frac{e^{-\tau \omega}}{1 + e^{-\omega}} = K(\tau,\omega),\]
this formula gives the desired result.
If $\omega < 0$, then \eqref{eq:k1} yields
\begin{equation} \label{eq:k4}
  K(-\tau^*,-\omega) = \frac{e^{-\tau^* \omega}}{1 + e^{\omega}}
\end{equation}
which is also numerically stable. We again have
\[\frac{e^{-\tau^* \omega}}{1 + e^{\omega}} = \frac{e^{(1-\tau)
\omega}}{1 + e^{\omega}} = K(\tau,\omega).\]

Let us illustrate the advantage with a concrete example. For simplicity,
we assume we are working in three-digit arithmetic. We first consider
the evaluation of $K(\tau,\omega)$ for $\tau = \num{0.501e-3}$ and
$\omega = 1000$. No problem arises in this case; we use the formula
\eqref{eq:k1}, and obtain the correct value
\[K(\tau,\omega) = \frac{e^{-0.501}}{1+e^{-1000}}.\]
However, if we instead want to calculate $K(\tau,\omega)$ for $\tau =
1-\num{0.501e-3} = 0.999499$ and $\omega = -1000$, the situation is
different.
In three-digit arithmetic, using the absolute format, we must round to $\tau = 0.999$.
Then we find
\[1-\tau = 0.001 \implies (1-\tau) \omega = -1,\]
and using \eqref{eq:k2} directly gives
\[K(\tau,\omega) = \frac{e^{-1}}{1+e^{-1000}},\]
which is far from the correct value.
If we instead use the relative format, $\tau$ is stored as
$\tau^* = \num{-0.501e-3}$, and using \eqref{eq:k4} gives precisely the
correct value.

In practice, with double precision arithmetic, using the absolute format
only leads to a significant loss of accuracy for very large
$\Lambda$ and small $\epsilon$. However, to enable calculations to high accuracy
in extreme scenarios, all subroutines in \texttt{libdlr} take in imaginary
time values in the relative format. Of course, maintaining full relative
precision also requires external procedures, such as those used to
evaluate Green's functions, to be similarly careful about this issue.

Many users are likely not operating in regimes in which it is important
to maintain full relative precision for extreme parameter values. These
users can simply ignore this discussion. However, they must
still convert imaginary time values to the relative format before using
them as inputs to \texttt{libdlr} subroutines (though, of course,
imaginary time values which are converted from the absolute format to
the relative format are only accurate to the original absolute
precision). The subroutine \texttt{abs2rel} performs this conversion.
Similarly, the subroutine \texttt{rel2abs} converts imaginary time
values in the relative format used by \texttt{libdlr} to the ordinary
absolute format.

\bibliographystyle{ieeetr}
{\footnotesize \bibliography{libdlr}}

\end{document}